\documentclass{article}

\PassOptionsToPackage{numbers}{natbib}

\usepackage[preprint]{neurips_2026}

\usepackage[utf8]{inputenc} 
\usepackage[T1]{fontenc}    
\usepackage{hyperref}       
\usepackage{url}            
\usepackage{booktabs}       
\usepackage{amsfonts}       
\usepackage{nicefrac}       
\usepackage{microtype}      
\usepackage[table]{xcolor}         

\usepackage{graphicx}
\usepackage{amsmath}
\usepackage{subcaption}
\usepackage{wrapfig} 
\usepackage{multirow}
\usepackage{subcaption}
\usepackage{tabularx}

\title{RipplePLM: Structural and Property Decoupling for Protein Mutation Effect Generation}

\author{%
  Liuzhenghao Lv$^{1,2,3}$ \quad
  Yuyang Liu$^{2,3,\dagger}$ \quad
  Yuyang Gao$^{1,2}$ \\
  \textbf{Li Yuan}$^{2,3}$ \quad
  \textbf{Yonghong Tian}$^{1,2,3,\dagger}$ \\[0.5em]
  \parbox{\textwidth}{\centering
    $^{1}$Beijing Key Laboratory of Brain-inspired Spiking Large Models,\\
    School of Computer Science, Peking University, Beijing, China\\[0.3em]
    $^{2}$Guangdong Provincial Key Laboratory of AI for Science
    and Computing (AI4SC),\\
    School of Artificial Intelligence for Science, Shenzhen Graduate School,\\
    Peking University, Shenzhen, China\\[0.3em]
    $^{3}$School of Electronic and Computer Engineering,
    Shenzhen Graduate School,\\
    Peking University, Shenzhen, China\\[0.5em]
    $^{\dagger}$Corresponding authors.\\
    Emails: \texttt{lvliuzh@stu.pku.edu.cn},
    \texttt{liuyuyang13@pku.edu.cn},
    \texttt{yhtian@pku.edu.cn}
  }
}

\begin{document}

\maketitle

\begin{abstract}
Protein mutation effect generation asks a model to describe the functional consequence of a point mutation in natural language. Existing protein-to-text systems typically encode mutation information into undifferentiated representations, overlooking the organization of mutation-induced evidence across structural and biochemical factors. We propose RipplePLM, a mutation-aware generation framework centered on Direct-Distal Cross-Attention (DDCA). By constructing a residue-level Mutation Perturbation Field from pre-trained protein language models, DDCA leverages predicted contact maps to organize mutation representations into two pathways: the mutation site's immediate contact neighborhood and its multi-hop distal context. To complement this structural decomposition, we further introduce the Property Latent Chain (PLChain), which injects expert-guided supervision of biochemical property changes (e.g., thermostability and optimal pH) into the LLM hidden-state pathway through latent property tokens. On MutaDescribe, RipplePLM improves over mutation-specific baselines on temporal and structural splits; under a matched-backbone comparison, average structural-split ROUGE-L increases from 22.23 to 35.65. Expert evaluation further shows a higher proportion of biologically accurate or relevant descriptions than the mutation-specific baseline. Additional ablations, representation diagnostics, and low-$N$ fitness regression experiments further support the effectiveness of the learned mutation-aware representations. Code: https://github.com/Lyu6PosHao/RipplePLM.
\end{abstract}

\section{Introduction}
\label{sec:intro}

Understanding the functional consequences of protein mutations is a central problem in protein engineering, disease variant interpretation, and biological discovery~\cite{meier2021language,brandes2023genome}. While many mutation-effect models focus on scalar prediction~\cite{hopf2017evmutation,riesselman2018deep,notin2023proteingym,gray2018quantitative}, experimental findings are often communicated as natural-language statements, such as whether a mutation weakens activity, alters stability, changes binding, or shifts an enzyme's preferred condition~\cite{kawabata1999protein,pires2015platinum,prevsern2023enzyme,mutaplm}.
These observations motivate \emph{mutation effect generation}, where the goal is to generate a natural-language description of the functional consequence of a given protein mutation~\cite{mutaplm}. This task is challenging because the model must ground fluent text in mutation-specific evidence. A useful description should therefore reflect not only the substituted residue itself, but also its structural neighborhood~\cite{guerois2002predicting,worth2011sdm,pires2014mcsm}, more distal contact context~\cite{hopf2017evmutation,li20223d}, and property-level cues~\cite{ramakrishnan2023understanding}. 
Rather than treating a mutation as a monolithic sequence change, a principled generation framework should reflect the underlying biophysical dependencies through which a substitution perturbs its immediate structural environment, propagates through the contact network, and gives rise to measurable biochemical changes that shape the functional outcome.

\begin{figure}[htbp]
    \centering
    \includegraphics[width=0.95\textwidth]{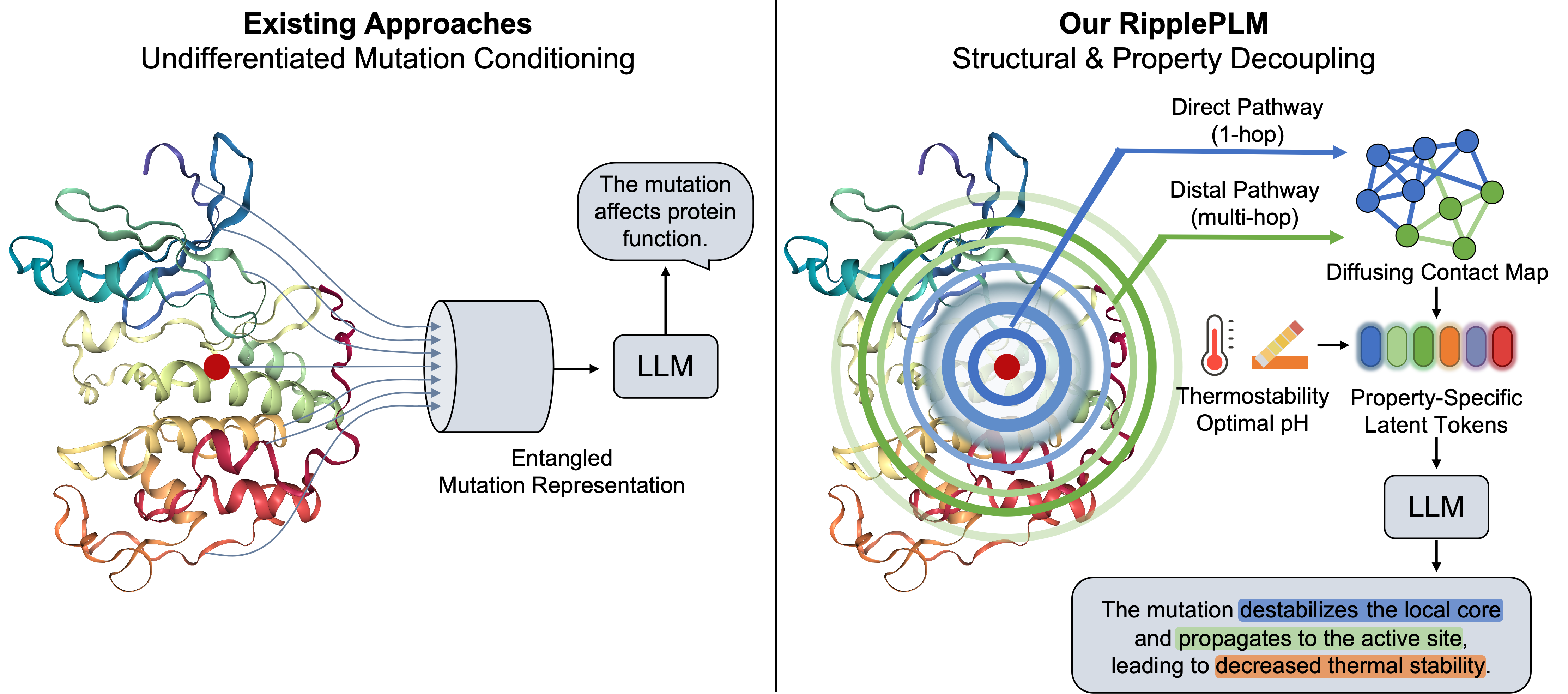}
    \caption{Comparison between existing mutation-effect generation approaches and RipplePLM. Existing methods encode mutation information into undifferentiated representations that obscure structural dependencies and property-level signals. RipplePLM instead organizes mutation representations through direct contact context, distal multi-hop context, and property-aware latent supervision, enabling more structured and biologically grounded generation.}
    \label{fig:motivation}
\end{figure}

As shown in Figure~\ref{fig:motivation}, existing protein--language systems~\cite{guo2023proteinchat,wang2024instructprotein,abdine2024prot2text,prollama,molinstructions,xu2023protst} typically align protein encoders with LLMs by projecting sequence or structure representations into the text model. 
Such interfaces are effective for broad protein captioning and instruction following, but mutation effect generation requires structured conditioning over mutation-specific evidence.
MutaPLM~\cite{mutaplm} moves closer to the target task by modeling protein deltas and using literature-derived reasoning supervision, yet its latent representations remain structurally undifferentiated and do not explicitly separate local contact evidence from distal contact context. In parallel, expert models for biochemical properties such as thermostability and optimal pH provide useful mutation-level signals~\cite{pptstab,ephod,cao2019deepddg,thumuluri2022netsolp}. Yet ordinary prompt text reduces these signals to coarse symbolic cues and leaves property supervision disconnected from the latent representations that drive generation~\cite{bengio2013representation,alain2016understanding}.

We propose RipplePLM, a mutation effect generation framework centered on Direct-Distal Cross-Attention (DDCA). RipplePLM first constructs a residue-level Mutation Perturbation Field~(MPF) by subtracting wild-type ESM-2 representations from mutant representations~\cite{lin2023evolutionary}. DDCA then uses the predicted contact map to produce two complementary mutation summaries: a \emph{direct} pathway that attends to the mutation site's contact neighborhood, and a \emph{distal} pathway that aggregates multi-hop contact-graph context. 
These summaries organize mutation-specific structural evidence into explicit direct and distal pathways before text generation.
Complementarily, we introduce Property Latent Chain (PLChain), which injects property-specific supervision into the LLM hidden-state pathway through expert-guided latent tokens for changes in thermostability and optimal pH.
Together, DDCA and PLChain implement structural and property decoupling at the representation level: DDCA organizes mutation evidence by contact relationships, while PLChain applies property-change supervision to the hidden states of property tokens.

We evaluate RipplePLM on the temporal and structural splits of MutaDescribe, including a matched-backbone MutaPLM reimplementation for a fairer comparison.
RipplePLM improves over evaluated mutation-specific baselines across the main generation metrics, while ablations and representation analyses support the effectiveness of both the structural decomposition and property-aware supervision mechanisms. Expert evaluation further shows a higher proportion of biologically accurate or relevant descriptions than reported for MutaPLM.
Our contributions are threefold: 
\begin{itemize}
    \item We introduce DDCA, a contact-guided mutation representation module that separates direct contact-neighborhood evidence from distal contact-graph context;
    \item We propose PLChain, an auxiliary latent property regularizer that integrates expert signals into the LLM hidden-state pathway;
    \item We provide a focused empirical study combining lexical and expert-based biological evaluations with ablations and representation diagnostics.
\end{itemize}

\section{Method}

\begin{figure}[t]
    \centering
    \includegraphics[width=\textwidth]{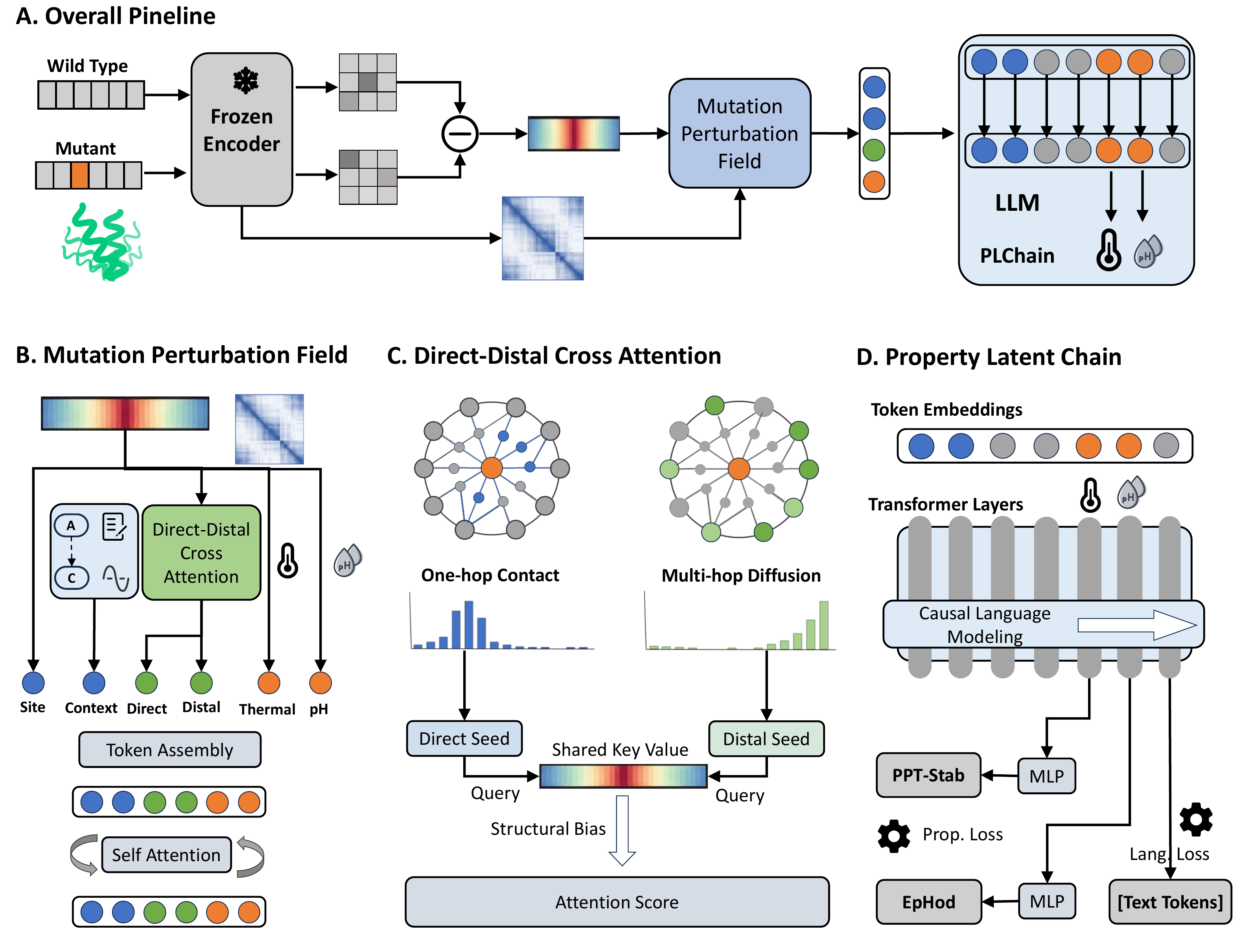}
    \caption{\textbf{Architecture of RipplePLM.} \textbf{(A)}~Pipeline overview: a frozen ESM-2 extracts the mutation perturbation field $\boldsymbol{\Delta}$, which MPF compresses into six conditioning tokens for the LLM. \textbf{(B)}~MPF assembles site, context, DDCA-derived structural tokens (direct/distal), and property queries, refined via self-attention. \textbf{(C)}~DDCA leverages predicted contact maps to organize mutation representations into direct (1-hop) and distal (multi-hop) cross-attention pathways using additive structural biases. \textbf{(D)}~PLChain regularizes latent property tokens with expert auxiliary losses while concurrently conditioning the autoregressive generation.}
    \label{fig:pipeline}
\end{figure}

Given a wild-type sequence $\mathbf{s}^{\text{wt}}$ of length $L$, a point mutation $m$ at position $i$, and optional textual context $c$ (the wild-type function description in our experiments), the task requires producing a description $\mathbf{y}$ of the mutation's functional consequences.
We denote the ESM-2 embedding dimension as $d_e$, the MPF hidden dimension as $d$, and the number of attention heads as $H$.
The overview of RipplePLM is shown in Figure~\ref{fig:pipeline}.

\subsection{Mutation Perturbation Field (MPF)}
\label{sec:mpf}

Our analysis (Section~\ref{sec:analysis}) shows that mutation-induced changes in ESM-2 representations extend beyond the mutation site and vary with distance on the predicted contact graph. We define MPF as a residue-indexed latent representation field formed by the differences between mutant and wild-type hidden states of a frozen ESM-2 encoder. DDCA organizes this field into direct (contact-level) and distal (multi-hop) components through contact-guided cross-attention, using predicted contact relationships as structural priors for mutation-effect generation.

\paragraph{Perturbation field.}
We encode $\mathbf{s}^{\text{wt}}$ and $\mathbf{s}^{\text{mt}}$ with a frozen ESM-2, extracting final-layer representations $\mathbf{H}^{\text{wt}}, \mathbf{H}^{\text{mt}} \in \mathbb{R}^{L \times d_e}$.
The perturbation field is:
\begin{equation}
    \boldsymbol{\Delta} = \mathbf{H}^{\text{mt}} - \mathbf{H}^{\text{wt}} \in \mathbb{R}^{L \times d_e},
\end{equation}
where $\mathbf{H}^{\text{mt}}$ is obtained by applying mutation $m$ to $\mathbf{s}^{\text{wt}}$.
We project $\boldsymbol{\Delta}$ through layer normalization and a linear layer to obtain $\hat{\boldsymbol{\Delta}} \in \mathbb{R}^{L \times d}$, and extract the site feature $\mathbf{z}_{\text{site}} = \hat{\boldsymbol{\Delta}}_i$.
Using the full field $\boldsymbol{\Delta}$ is important because ESM-2 contextualizes each residue by the whole sequence.
Changing one amino acid can therefore change hidden states away from the mutated position.
Mean pooling this field would remove information about where the perturbation appears, while using only $\boldsymbol{\Delta}_i$ would discard non-local changes.
DDCA keeps the field spatially indexed until the contact-guided aggregation step.

\paragraph{Direct-Distal Cross-Attention (DDCA).}
We leverage the contact map $\mathbf{C} \in [0,1]^{L \times L}$ predicted from the wild-type sequence $\mathbf{s}^{\text{wt}}$ by the frozen ESM-2 contact head. ESM-2 has been shown to capture structural contacts with high accuracy from sequence alone~\cite{rao2021contact,lin2023evolutionary}, thereby avoiding dependence on external structure prediction at inference. The wild-type map provides a shared pre-mutation reference topology for the direct and distal pathways to organize mutation-induced representation changes.
Let $\mathbf{m} \in \{0,1\}^{L}$ be the non-padding mask and $\tilde{\mathbf{C}} = \mathbf{C} \odot (\mathbf{m}\mathbf{m}^{\top})$ the masked contact matrix; let $\mathbf{v}_i$ be a binary mask that excludes position $i$ itself.
Let $\operatorname{Norm}(\mathbf{x})=\mathbf{x}/(\sum_j x_j+\epsilon)$ with a small $\epsilon$ for empty neighborhoods.
The \emph{direct} pathway weights are obtained by $\ell_1$-normalizing the mutation site's contact row:
\begin{equation}
    \mathbf{w}^{\text{dir}} = \operatorname{Norm}\!\left(\tilde{\mathbf{C}}_{i,:} \odot \mathbf{v}_i\right).
\end{equation}
For the \emph{distal} pathway, we define a structure diffusion kernel with learnable per-head decay $\alpha_h = \sigma(\tilde{\alpha}_h)$ (where $\sigma$ denotes the sigmoid function) over the row-normalized contact matrix $\bar{\mathbf{C}}$:
\begin{equation}
    \mathbf{S}_h = \sum_{k=1}^{K} \alpha_h^{k} \cdot (\bar{\mathbf{C}}^{k})_{i,:},
\end{equation}
where $K$ is the diffusion horizon.
The distal weights retain multi-hop ($k \geq 2$) contributions:
\begin{equation}
    \mathbf{w}^{\text{dist}}_h = \operatorname{Norm}\!\left((\mathbf{S}_h - \alpha_h \cdot \bar{\mathbf{C}}_{i,:}) \odot \mathbf{v}_i\right), \quad h = 1, \ldots, H.
\end{equation}

The direct and distal weights play different roles.
The direct weights identify the first structural shell around the mutation site, which is the most immediate source of contact-level evidence.
The distal weights use powers of the row-normalized contact matrix to reach residues connected through multiple structural steps.
The per-head decay $\alpha_h$ lets different attention heads prefer different effective radii instead of fixing a single hop threshold for all proteins and mutations.
Subtracting the first-hop term from $\mathbf{S}_h$ retains contributions from paths of length two or greater, giving the distal pathway a multi-hop structural prior.

DDCA performs two parallel cross-attention operations over $\hat{\boldsymbol{\Delta}}$.
Each pathway first computes a seed feature $\mathbf{p}$ by weighted-averaging $\hat{\boldsymbol{\Delta}}$ with its structural weights (direct uses $\mathbf{w}^{\text{dir}}$; distal averages across heads):
\begin{equation}
    \mathbf{p}^{\text{dir}} = \sum\nolimits_{j} w^{\text{dir}}_j \hat{\boldsymbol{\Delta}}_j, \quad
    \mathbf{p}^{\text{dist}} = \tfrac{1}{H}\sum\nolimits_{h}\sum\nolimits_{j} w^{\text{dist}}_{h,j} \hat{\boldsymbol{\Delta}}_j.
\end{equation}
These seeds are projected into pathway-specific queries $\mathbf{q}$, which attend over shared keys/values derived from $\hat{\boldsymbol{\Delta}}$.
The structural weights enter as learnable additive biases on the attention logits, so that each pathway's attention is guided by---but not restricted to---its structural prior:
\begin{equation}
\begin{aligned}
a^{\text{dir}}_{h,j} &= \frac{\mathbf{q}^{\text{dir}}_h \cdot \mathbf{k}_{h,j}}{\sqrt{d/H}} + \sigma(\beta^{\text{dir}}_h)\, w^{\text{dir}}_j, \quad &
a^{\text{dist}}_{h,j} &= \frac{\mathbf{q}^{\text{dist}}_h \cdot \mathbf{k}_{h,j}}{\sqrt{d/H}} + \sigma(\beta^{\text{dist}}_h)\, w^{\text{dist}}_{h,j},
\end{aligned}
\end{equation}
where $\beta^{\text{dir}}_h, \beta^{\text{dist}}_h$ are learnable scalars controlling the strength of the structural bias.
Each pathway produces a pooled vector $\mathbf{o}^{\text{dir}}, \mathbf{o}^{\text{dist}} \in \mathbb{R}^{d}$.
We use additive biases rather than hard structural masks because predicted contacts are noisy and because useful mutation evidence need not be limited to residues selected by the prior.
The bias initializes each pathway with a different structural preference, while the content-based attention term can upweight or downweight residues according to the learned generation objective.

During both training stages, we apply a pathway-overlap regularizer to encourage complementary direct and distal structural priors; its definition and the complete training objectives are given in Appendix~\ref{app:overlap}.

\paragraph{Token assembly.}
MPF assembles six tokens: $\mathbf{t}_1$ (site, from $\mathbf{z}_{\text{site}}$), $\mathbf{t}_2$ (direct, from $\mathbf{o}^{\text{dir}}$), $\mathbf{t}_3$ (distal, from $\mathbf{o}^{\text{dist}}$), $\mathbf{t}_4$ (context, from a mutation-type embedding and positional encoding), and two learnable property queries $\mathbf{t}_5, \mathbf{t}_6$ for PLChain.
These are stacked into $\mathbf{T}_0 \in \mathbb{R}^{6 \times d}$ and refined by self-attention:
\begin{equation}
    \mathbf{T} = \text{LayerNorm}\!\left(\mathbf{T}_0 + \text{Attention}(\mathbf{T}_0, \mathbf{T}_0, \mathbf{T}_0)\right) \in \mathbb{R}^{6 \times d}.
\end{equation}

\subsection{Property Latent Chain (PLChain)}
\label{sec:plchain}

PLChain introduces a lightweight property-aware pathway for grounding generation in biochemical signals beyond structural perturbation. We focus on thermostability and pH optimum, which characterize the environmental conditions governing protein structure and function: thermostability reflects the ability to retain an active conformation under thermal stress~\cite{haki2003developments}, while pH optimum is closely associated with catalytic activity and stability~\cite{talley2010ph,bisswanger2014enzyme}. Modeling changes in these properties aligns auxiliary supervision with the mutation-effect generation objective. Specifically, we construct pseudo-labels for changes in thermostability and pH optimum using PPT-Stab~\cite{pptstab} and EpHod~\cite{ephod}, respectively. These labels guide the hidden states of \emph{expert-guided latent nodes} within the LLM during training.

\paragraph{Token injection.}
Special tokens \texttt{<|protein\_token|>}, \texttt{<|thermo\_token|>}, and \texttt{<|ph\_token|>} are added to the LLM vocabulary.
During the forward pass, the four structural tokens $\hat{\mathbf{t}}_{1\text{--}4}$ replace repeated \texttt{<|protein\_token|>} embeddings, while property queries $\hat{\mathbf{t}}_{5\text{--}6}$ replace \texttt{<|thermo\_token|>} and \texttt{<|ph\_token|>} via masked scattering.

\paragraph{Expert-guided auxiliary supervision.}
After the LLM forward pass, lightweight auxiliary heads project the property token hidden states onto expert label spaces:
\begin{equation}
    \hat{y}^{\text{thermo}} = \text{MLP}^{\text{thermo}}\!\left(\mathbf{h}_{\texttt{thermo}}\right), \quad
    \hat{y}^{\text{ph}} = \text{MLP}^{\text{ph}}\!\left(\mathbf{h}_{\texttt{ph}}\right),
\end{equation}
aligned with expert pseudo-labels via cross-entropy.
The same hidden states also flow into autoregressive generation, so the auxiliary loss acts as a {representation regularizer}: property tokens are encouraged to encode biochemically grounded information while remaining optimized for generation.

\subsection{Training Procedure}
\label{sec:training}

RipplePLM is trained in two stages with ESM-2 frozen throughout; this staged approach first learns a structural perturbation interface before introducing the additional property supervision signal, reducing early interference between the auxiliary and language modeling objectives.
\textbf{Stage~1} trains MPF under the language modeling loss $\mathcal{L}_{\text{LM}}$; property heads are inactive.
\textbf{Stage~2} activates PLChain with LoRA ($r{=}64$) across all projections, optimizing the task objective:
\begin{equation}
    \mathcal{L}_{\text{task}}(t) = \bigl(1 - w_{\text{prop}}(t)\bigr)\mathcal{L}_{\text{LM}} + w_{\text{prop}}(t)\mathcal{L}_{\text{prop}},
\end{equation}
where $w_{\text{prop}}(t)$ is annealed from $0.4$ to $0.2$.
This schedule lets expert supervision shape the property token representations early in training, then gradually gives more weight to the language modeling objective so that generation quality is not dominated by the auxiliary task in later stages.

\section{Experiments}

\subsection{Experimental Setup}

\paragraph{Dataset and metrics.}

We evaluate on MutaDescribe~\cite{mutaplm}, a benchmark providing wild-type sequences, point mutations, and functional text descriptions. We test generalization using two splits: the \textit{temporal split} (chronological partitioning to mitigate information leakage) and the \textit{structural split} (stratified by structural distance to the training set, ranging from \textit{Easy} to \textit{Hard}, to assess robustness against distribution shifts). Following prior work~\cite{mutaplm}, we evaluate generation quality using BLEU-2, BLEU-4~\cite{papineni2002bleu}, ROUGE-1, ROUGE-2, ROUGE-L~\cite{lin2004rouge}, and METEOR~\cite{banerjee2005meteor}. Since these metrics capture different textual properties (e.g., exact $n$-gram matching vs. semantic recall), we report the full suite on the temporal split, while adopting ROUGE-L and BLEU-2 as summary metrics for the structural split.

\paragraph{Baselines.}

We benchmark our framework against four categories of baselines: (1) general protein--language models (ProLLaMA~\cite{prollama}, Galactica-6.7B~\cite{galactica}, Mol-Instructions~\cite{molinstructions}); (2) prompted LLMs (GPT-4-0613~\cite{gpt4}); (3) protein-conditioned generation frameworks (AugmentedESM~\cite{augmentedesm}, fine-tuned ESM-2~\cite{lin2023evolutionary}, GPT-4 + ESM-2, GPT-4 + OntoProtein~\cite{ontoprotein}); and (4) the state-of-the-art mutation-specific generation model, MutaPLM~\cite{mutaplm}. To control for LLM backbone choice, we introduce {MutaPLM$^{*}$}, a retrained version of MutaPLM using the same base LLM as RipplePLM. For a fair comparison, we use the authors' official implementation and released hyperparameters. This setup enables a controlled comparison between the two frameworks under the same LLM backbone.

\begin{table}[t]
\centering
\caption{Performance comparison on the \textbf{temporal split}. Best results are in \textbf{bold}. B/R denote BLEU/ROUGE. *: same base LLM as RipplePLM.}
\label{tab:temporal}
\small
\resizebox{0.9\linewidth}{!}{%
\begin{tabular}{@{}llcccccc@{}}
\toprule
Type & Model & B-2$\uparrow$ & B-4$\uparrow$ & METEOR$\uparrow$ & R-1$\uparrow$ & R-2$\uparrow$ & R-L$\uparrow$ \\
\midrule
\multirow{3}{*}{\textit{General}} 
& ProLLaMA             & 0.69  & 0.21 & 3.33  & 0.83  & 0.04  & 0.80  \\
& Galactica-6.7B       & 3.50  & 1.31 & 5.61  & 7.44  & 0.85  & 6.17  \\
& Mol-Instructions     & 0.58  & 0.08 & 4.90  & 5.41  & 0.13  & 4.55  \\
\midrule
\multirow{3}{*}{\textit{Conditioned}} 
& GPT-4 (5-shot, $k$NN)   & 9.30  & 4.25 & 15.08 & 13.92 & 2.29  & 11.84 \\
& AugmentedESM         & 7.00  & 3.12 & 11.29 & 12.03 & 2.84  & 10.12 \\
& Fine-tuned ESM-2     & 6.90  & 3.83 & 13.86 & 14.21 & 4.63  & 12.62 \\
\midrule
\multirow{3}{*}{\textit{Mutation-spec.}} 
& MutaPLM              & 10.83 & 6.15 & 17.84 & 18.99 & 6.92  & 16.51 \\
& MutaPLM$^{*}$        & 9.44  & 3.93 & 15.28 & 21.45 & 7.93  & 18.22 \\
\rowcolor{gray!10}
& \textbf{RipplePLM}   & \textbf{13.41} & \textbf{6.68} & \textbf{25.23} & \textbf{31.15} & \textbf{16.00} & \textbf{28.56} \\
\bottomrule
\end{tabular}%
}
\end{table}

\paragraph{Implementation.}
RipplePLM employs a frozen ESM-2 (\texttt{esm2\_t30\_150M\_UR50D}) as the foundational protein encoder, paired with DeepSeek-R1-Distill-Llama-8B~\cite{guo2025deepseek} as the autoregressive LLM backbone, adapted via LoRA~\cite{hu2022lora}. Within the DDCA module, we configure $H{=}8$ attention heads and a structural diffusion horizon of $K{=}4$.
Implementation details, including backbone selection rationale, dataset configuration, training hyperparameters, optimization settings, and inference protocols, are provided in Appendix~\ref{app:impl}.

\subsection{Main Results}

\paragraph{Temporal split.}
Table~\ref{tab:temporal} details the performance on the temporal split, which explicitly tests generalization to newly documented mutations. RipplePLM consistently outperforms all evaluated baselines across all six metrics. Notably, compared to the previous state-of-the-art MutaPLM, it yields substantial absolute improvements (e.g., $+7.39$ METEOR, $+12.05$ ROUGE-L). Since METEOR accounts for synonym matches, the gain suggests improved alignment with reference descriptions beyond exact-word matching. Furthermore, against the matched-backbone MutaPLM$^{*}$, RipplePLM still increases ROUGE-L from $18.22$ to $28.56$, supporting the effectiveness of our framework under the same LLM backbone.

\paragraph{Structural split.}
Table~\ref{tab:structural} details performance across varying degrees of structural novelty, serving as a rigorous stress-test for out-of-distribution (OOD) generalization. While baseline models suffer severe performance degradation as test proteins deviate structurally from the training manifold, RipplePLM maintains robust generation fidelity, achieving a striking $+13.42$ absolute gain in average ROUGE-L over the matched-backbone MutaPLM$^{*}$. More importantly, RipplePLM demonstrates exceptional structural resilience: its performance on the \textit{Hard} subset (ROUGE-L 31.27) surpasses the previous state-of-the-art MutaPLM's performance on the \textit{Easy} subset (ROUGE-L 25.80). This cross-difficulty comparison supports the ability of DDCA's contact-guided representations to generalize across varying degrees of structural similarity to the training set.

\paragraph{Additional evaluation.}
Expert evaluation yields a higher proportion of biologically accurate or relevant descriptions for RipplePLM than reported for MutaPLM (48.56\% vs.\ 40.06\%; Table~\ref{tab:expert_eval}). RipplePLM also remains ahead of MutaPLM across all six temporal metrics after truncating over-length outputs to their reference lengths (Table~\ref{tab:length_control}). Bootstrap confidence intervals characterize the sampling uncertainty of the reported scores (Table~\ref{tab:bootstrap_ci}). Recent general-purpose LLMs evaluated in a one-shot setting remain well below RipplePLM on the structural split (Table~\ref{tab:recent_llms}), highlighting the importance of mutation-specific modeling for this task. Detailed protocols and results are provided in Appendix~\ref{app:additional_evaluation}.

\paragraph{Qualitative case studies.}
Beyond aggregate metrics, qualitative examples further illustrate the advantages of structured mutation conditioning. As shown in Appendix~\ref{app:case}, RipplePLM more consistently preserves mutation-specific mechanistic evidence, including quantitative biochemical changes, interaction targets, and substrate-specific functional effects. For example, on the hard structural-split mutation \textit{N226A}, RipplePLM correctly captures the ``100-fold decrease in GTPase activity,'' whereas MutaPLM$^{*}$ collapses the description into the less specific statement ``abolishes GTP hydrolysis.'' Similarly, for \textit{G93A}, RipplePLM retains the TDIF--TDR interaction context instead of drifting toward a broader developmental phenotype. We also include competitive counterexamples where both models remain imperfect, highlighting realistic failure modes such as missing secondary clauses or incomplete assay-specific details.

\begin{table}[t]
\centering
\caption{Performance comparison on the \textbf{structural split}. R-L denotes ROUGE-L and BL-2 denotes BLEU-2. Best results are in \textbf{bold}. *: same base LLM as RipplePLM.}
\label{tab:structural}
\small
\resizebox{0.9\linewidth}{!}{%
\begin{tabular}{@{}l cc cc cc cc@{}}
\toprule
 & \multicolumn{2}{c}{Easy} & \multicolumn{2}{c}{Medium} & \multicolumn{2}{c}{Hard} & \multicolumn{2}{c}{Average} \\
\cmidrule(lr){2-3} \cmidrule(lr){4-5} \cmidrule(lr){6-7} \cmidrule(lr){8-9}
Model & R-L & BL-2 & R-L & BL-2 & R-L & BL-2 & R-L & BL-2 \\
\midrule
ProLLaMA             & 1.02  & 0.64  & 1.00  & 0.91  & 1.03  & 0.70  & 1.02  & 0.74  \\
Mol-Instructions     & 5.10  & 0.65  & 5.19  & 0.65  & 5.56  & 0.90  & 5.27  & 0.73  \\
Galactica-7B         & 6.53  & 3.52  & 7.64  & 3.58  & 7.33  & 2.88  & 7.13  & 3.33  \\
\midrule
GPT-4 (1-shot)       & 8.04  & 2.93  & 9.96  & 3.42  & 9.62  & 2.69  & 9.14  & 3.00  \\
GPT-4 (5-shot)        & 10.46 & 2.51  & 10.31 & 2.81  & 10.79 & 1.88  & 10.52 & 2.40  \\
GPT-4 (5-shot, $k$NN) & 11.63 & 9.63 & 12.98 & 10.88 & 12.46 & 8.63 & 12.31&  9.69   \\
GPT-4 + ESM-2        & 11.69 & 11.09 & 13.02 & 11.50 & 12.77 & 8.48  & 12.45 & 10.37 \\
GPT-4 + OntoProtein  & 11.84 & 10.93 & 12.69 & 11.22 & 12.81 & 8.17  & 12.42 & 10.13 \\
\midrule
AugmentedESM         & 11.60 & 8.33  & 11.40 & 7.46  & 10.73 & 6.95  & 11.26 & 7.62  \\
Fine-tuned ESM-2     & 20.49 & 9.37  & 11.87 & 5.95  & 11.34 & 3.32  & 14.88 & 6.36  \\
\midrule
MutaPLM              & 25.80 & 18.77 & 21.07 & 12.59 & 16.51 & 8.69  & 21.34 & 13.61 \\
MutaPLM$^{*}$        & 24.91 & 11.67 & 23.71 & 12.13 & 18.80 & 8.17  & 22.23 & 10.66 \\
\rowcolor{gray!10}
\textbf{RipplePLM}   & \textbf{42.91} & \textbf{22.57} & \textbf{31.56} & \textbf{12.67} & \textbf{31.27} & \textbf{11.91} & \textbf{35.65} & \textbf{15.78} \\
\bottomrule
\end{tabular}%
}
\end{table}


\subsection{Ablation Studies}
\label{sec:ablation}

\begin{wraptable}{r}{0.55\textwidth}
\centering
\vspace{-1em}
\small
\caption{\textbf{Ablation study on the temporal split.} Replacing PLChain with discrete prompt labels or removing architectural components consistently degrades performance.}
\label{tab:ablation}
\begin{tabular}{@{}lcc@{}}
\toprule
Variant & BLEU-2$\uparrow$ & ROUGE-L$\uparrow$ \\
\midrule
\textbf{RipplePLM (Full)} & \textbf{13.41} & \textbf{28.56} \\
\midrule
w/o Distal Pathway      & 12.02 & 25.38 \\
Discrete Labels & 11.70 & 25.06 \\
w/o PLChain             & 11.35 & 24.66 \\
\bottomrule
\end{tabular}
\vspace{-1em}
\end{wraptable}

We first ablate the core DDCA design and the auxiliary PLChain signal on the temporal split to isolate their contributions (Table~\ref{tab:ablation}). Removing the distal pathway from DDCA degrades BLEU-2 by $1.39$ and ROUGE-L by $3.18$, validating the necessity of capturing multi-hop structural context beyond the immediate contact shell. Removing PLChain incurs a drop of $2.06$ in BLEU-2 and $3.90$ in ROUGE-L, supporting the added value of continuous property supervision. 

To rigorously trace the source of this property-aware gain, we introduce two controls: training with shuffled property pseudo-labels yields negligible recovery over the no-PLChain variant, while replacing the latent tokens with discrete text labels in the LLM prompt strictly underperforms the full model. This confirms that PLChain derives its strength from continuous, expert-guided latent regularization rather than mere token capacity or explicit text conditioning.

\paragraph{Function context and encoder size.}
Removing the wild-type function description reduces RipplePLM's structural-split ROUGE-L from 35.65 to 28.55, retaining 80.1\% of its score and exceeding the published no-function MutaPLM result by 10.01 points (Table~\ref{tab:function_context}). Replacing the frozen ESM-2 150M encoder with its 650M counterpart improves four of six metrics under the same training settings, with ROUGE-L increasing to 36.43 (Table~\ref{tab:encoder_size}). These experiments examine the contribution of function context and protein encoder capacity; details are provided in Appendix~\ref{app:additional_controls}.


\begin{figure}[t]
    \centering
    \begin{subfigure}[b]{0.48\textwidth}
        \centering
        \includegraphics[width=0.9\textwidth]{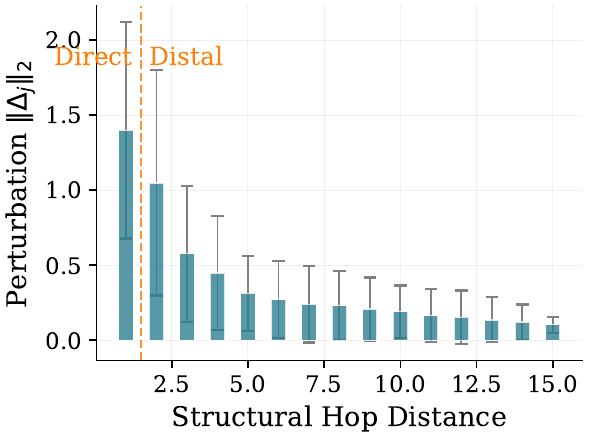} 
        \caption{Structural Hop Decay}
        \label{fig:diag_hop}
    \end{subfigure}
    \hfill
    \begin{subfigure}[b]{0.48\textwidth}
        \centering
        \includegraphics[width=0.9\textwidth]{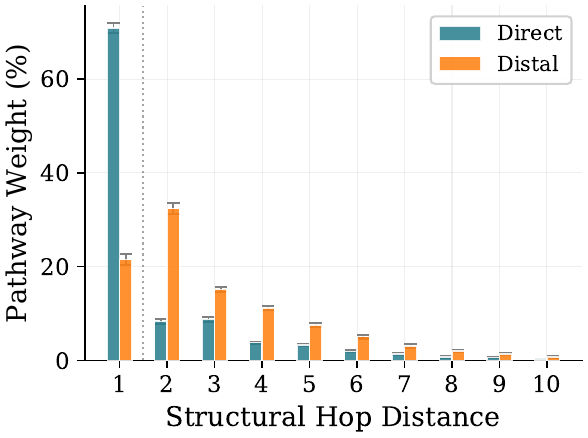}
        \caption{DDCA Structural Weight Distribution}
        \label{fig:diag_attn}
    \end{subfigure}
    
    \vspace{1em} 
    
    \begin{subfigure}[b]{0.48\textwidth}
        \centering
        \includegraphics[width=0.75\textwidth]{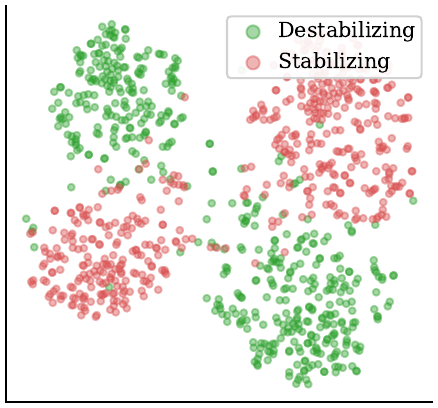}
        \caption{Thermostability (t-SNE)}
        \label{fig:diag_thermo}
    \end{subfigure}
    \hfill
    \begin{subfigure}[b]{0.48\textwidth}
        \centering
        \includegraphics[width=0.75\textwidth]{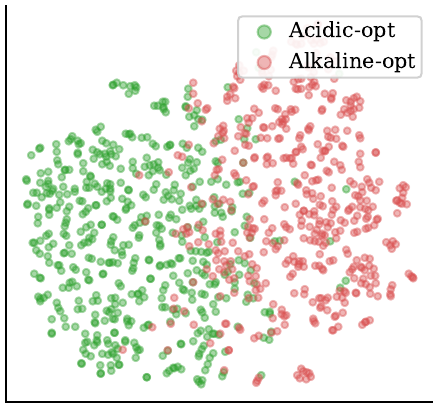}
        \caption{Optimal pH (t-SNE)}
        \label{fig:diag_ph}
    \end{subfigure}
    
    \caption{\textbf{Representation Diagnostics.} \textbf{(a)} Mutation perturbation magnitude remains significantly above background across multiple structural hops. \textbf{(b)} DDCA structural pathway weights cover different contact neighborhoods: the direct pathway concentrates on 1-hop residues, while the distal pathway assigns the most weight to 2-hop residues. \textbf{(c, d)} t-SNE visualizations of PLChain latent tokens after LLM processing show clear, biologically meaningful clustering for thermostability and optimal pH.}
    \label{fig:diagnostics}
\end{figure}

\subsection{Representation Analysis}
\label{sec:analysis}

We conduct representational diagnostics to examine whether the learned intermediate states reflect the intended structural and biochemical roles of DDCA and PLChain (Figure~\ref{fig:diagnostics}).

\paragraph{Perturbation decay.} 
We first evaluate the spatial scope of the ESM-2 perturbation field. Across 200 sampled temporal-test proteins, the per-residue perturbation $\|\mathbf{h}^{\text{mt}}_j-\mathbf{h}^{\text{wt}}_j\|_2$ peaks at the mutation site and decays smoothly. Notably, the near-site signal ($\leq 5$ residues) is $10.77\times$ the background noise ($> 100$ residues). On the predicted contact graph, this signal remains $9.1\times$ above background at 2 hops and $5.0\times$ at 3 hops (Figure~\ref{fig:diagnostics}a). These elevated representation differences at multi-hop distances motivate a dedicated distal pathway to capture mutation-induced changes beyond direct contacts.

\paragraph{DDCA structural weights and feature separation.}
The structural pathway weights exhibit different spatial coverage consistent with the DDCA design (Figure~\ref{fig:diagnostics}b). Across the analyzed temporal-test examples, the direct pathway assigns 70.9\% of its normalized weight to 1-hop residues, while the distal pathway assigns 21.5\% to 1-hop residues and 58.7\% to 2--4 hop residues, peaking at 2 hops. These statistics describe the contact-derived structural priors; the sampling and normalization protocol is given in Appendix~\ref{app:pathway_diagnostics}. Separately, direct and distal token representations have a silhouette score of $0.74$, indicating separation in feature space.

\paragraph{Property-token specificity.} 
For PLChain, we verify that the latent property tokens actively encode the targeted expert knowledge. Matched linear probes trained on the LLM-output property tokens achieve remarkable accuracies of \textbf{90.2\%} for thermostability and \textbf{88.3\%} for optimal pH. Visualized via t-SNE (Figure~\ref{fig:diagnostics}c, d), these representations form highly distinct, property-specific clusters (e.g., stabilizing vs. destabilizing mutations). Crucially, the pre-LLM MPF queries hover near random chance ($\sim$55\%), demonstrating that property specificity is not trivially embedded in the initial queries, but dynamically formed during the LLM's forward pass guided by the auxiliary loss.

\paragraph{Predicted-contact confidence.}
The proportion of biologically accurate or relevant descriptions increases from 44.23\% to 47.12\% and 54.22\% across low-, middle-, and high-confidence groups, respectively (Table~\ref{tab:contact_confidence}). This association links generation performance to model-reported contact confidence; the grouping protocol is described in Appendix~\ref{app:contact_confidence}.

\subsection{Low-N Fitness Regression}
\label{sec:fitness_main}

\begin{wraptable}{r}{0.55\textwidth}
\centering
\vspace{-1em}
\small
\caption{\textbf{Low-$\boldsymbol{N}$ fitness regression} (Spearman $\rho$, mean$\pm$std over 5 runs). Best mean is in \textbf{bold}.}
\label{tab:fitness_main}
\begin{tabular}{@{}lcc@{}}
\toprule
Model & Spike-ACE2 & avGFP \\
\midrule
Ridge Regression     & 0.335$\pm$0.052 & 0.298$\pm$0.071 \\
ESM-2                & 0.331$\pm$0.041 & 0.554$\pm$0.013 \\
Augmented ESM        & 0.363$\pm$0.021 & 0.497$\pm$0.096 \\
Augmented EVmutation & 0.354$\pm$0.044 & 0.512$\pm$0.034 \\
ConFit               & 0.412$\pm$0.033 & 0.564$\pm$0.035 \\
Tranception\_L       & 0.488$\pm$0.040 & 0.594$\pm$0.019 \\
MutaPLM              & 0.481$\pm$0.028 & 0.596$\pm$0.032 \\
\rowcolor{gray!10}
\textbf{RipplePLM}   & \textbf{0.491$\pm$0.061} & \textbf{0.598$\pm$0.067} \\
\bottomrule
\end{tabular}
\vspace{-1em}
\end{wraptable}

To ascertain whether MPF representations capture generalized biological information beyond text generation, we follow the MutaPLM protocol~\cite{mutaplm} for low-$N$ fitness regression. Using only 192 labeled single-point mutations for training, we train a lightweight regressor on frozen protein features without any textual supervision. We evaluate on two contrasting ProteinGym datasets: Spike-ACE2 (binding phenotype on a long viral protein) and avGFP (fluorescence on a shorter protein). 

As detailed in Table~\ref{tab:fitness_main}, RipplePLM achieves the highest mean Spearman correlation on both datasets. The standard deviations overlap with those of the strongest baselines, and the results suggest that the learned mutation representations provide useful features for fitness regression with limited labeled data.

\section{Related Work}
\paragraph{Protein--language alignment.}
Protein--language alignment aims to connect protein representations with natural-language descriptions, biomedical knowledge, or instruction-following capabilities~\cite{guo2023proteinchat,wang2024instructprotein,abdine2024prot2text,prollama,molinstructions,xu2023protst,ontoprotein,lv2026navigating,xu2023multilingual,li2025detect,liu2024prott3,zhuo2024protllm}. Early approaches such as ProtST~\cite{xu2023protst}, OntoProtein~\cite{ontoprotein}, and BioTranslator~\cite{xu2023multilingual} align proteins with text or ontology annotations through contrastive or knowledge-guided pretraining, while Prot2Text~\cite{abdine2024prot2text}, ProtT3~\cite{liu2024prott3}, and related protein-to-text systems demonstrate protein-conditioned description generation. More recent LLM-based systems, including ProteinChat~\cite{guo2023proteinchat}, InstructProtein~\cite{wang2024instructprotein}, ProLLaMA~\cite{prollama}, Mol-Instructions~\cite{molinstructions}, and ProtLLM~\cite{zhuo2024protllm}, project protein representations into general-purpose language models for protein-aware dialogue and instruction following. Although these methods establish the feasibility of protein-conditioned generation, most treat the protein input as a static object. Mutation effect generation instead requires localized and mutation-aware conditioning over wild-type/mutant differences. AugmentedESM~\cite{augmentedesm} and ESM-based generation baselines provide protein-conditioned generation signals, but they do not explicitly separate direct contact-neighborhood evidence from distal structural context. RipplePLM instead introduces contact-guided mutation perturbation representations for structured mutation conditioning.

\paragraph{Mutation effect prediction and generation.}
Mutation effect prediction has been extensively studied through evolutionary, probabilistic, structure-aware, and protein-language-model-based approaches~\cite{meier2021language,hopf2017evmutation,riesselman2018deep,notin2023proteingym,laine2019gemme,frazer2021eve,notin2022tranception,marquet2022embeddings,marquet2024expert,nijkamp2023progen2,cheng2023alphamissense,cheng2024zero}. Representative methods include EVmutation~\cite{hopf2017evmutation}, DeepSequence~\cite{riesselman2018deep}, GEMME~\cite{laine2019gemme}, EVE~\cite{frazer2021eve}, ESM-1v~\cite{meier2021language}, Tranception~\cite{notin2022tranception}, ProGen2~\cite{nijkamp2023progen2}, AlphaMissense~\cite{cheng2023alphamissense}, and ProMEP~\cite{cheng2024zero}. Other methods specialize in biochemical properties such as stability, thermostability, solubility, and enzyme optimal pH~\cite{guerois2002predicting,pires2014mcsm,pptstab,ephod,cao2019deepddg,thumuluri2022netsolp,kellogg2011role,li2020predicting,blaabjerg2023rapid,dieckhaus2024thermompnn}. However, these systems typically output scalar scores or class labels rather than natural-language explanations. MutaPLM~\cite{mutaplm} formalizes mutation effect generation with a protein delta network and literature-mined chain-of-thought-style supervision~\cite{wei2022chain,kojima2022large}. RipplePLM differs primarily in its representation decoupling strategy: DDCA organizes mutation representations into direct contact-neighborhood and distal contact-graph pathways before generation, while PLChain introduces auxiliary property-aware latent supervision.
\section{Conclusion}

We presented RipplePLM, a mutation effect generation framework that organizes mutation evidence through structured conditioning rather than undifferentiated protein representations. At its core, DDCA separates mutation-induced signals into direct contact-neighborhood evidence and distal contact-graph context, while PLChain introduces property-aware latent supervision to incorporate biochemical signals into the generation pathway. Experiments on MutaDescribe show consistent improvements over evaluated mutation-specific baselines on both temporal and structural splits, complemented by expert assessment of biological relevance. Low-$N$ fitness regression further suggests that MPF representations provide useful features for fitness prediction. Future work will extend RipplePLM toward multi-mutation settings and larger community-curated benchmarks that connect mutation descriptions with richer experimental contexts.

\paragraph{Limitations.}
RipplePLM focuses on single-point mutations, consistent with current mutation-description benchmarks; extending effect generation to multi-mutation settings will require broader community efforts to curate datasets with epistatic annotations. DDCA remains sequence-only and efficient by using ESM-2-predicted contact maps, though its structural signal may depend on contact prediction quality. Performance also depends in part on the availability and quality of wild-type function annotations. Our representation analyses characterize learned mutation signals; linking these signals to biophysical mechanisms requires experimental validation.

\paragraph{Broader Impacts.}
RipplePLM may assist protein engineering, variant interpretation, and literature-driven biological discovery by generating structured descriptions of mutation effects. Its outputs should be used for hypothesis generation and decision support rather than as substitutes for experimental validation or clinical expertise.

\begin{ack}
This work was supported by
the Guangdong Grants (Grant No. 2023ZT10X075),
the National Natural Science Foundation of China (No. 62425101, No. 62606021),
and the National Key R\&D Program of China (No. 2026ZD0127200).
\end{ack}

\clearpage

\bibliographystyle{unsrtnat}   
\bibliography{references}      


\appendix

\newpage
\appendix

\section{Implementation Details}
\label{app:impl}

\paragraph{Model and infrastructure.}
We use ESM-2~(esm2\_t30\_150M\_UR50D)~\cite{lin2023evolutionary} as the frozen protein encoder and initialize the language decoder from DeepSeek-R1-Distill-Llama-8B. DeepSeek-R1-Distill-Llama-8B is a reasoning-oriented dense language model obtained by distilling the reasoning behaviors of DeepSeek-R1 into the Llama architecture. Unlike standard instruction-tuned LLMs optimized primarily for conversational fluency, DeepSeek-R1-style models are trained to preserve intermediate reasoning patterns during generation.

We adopt this backbone because mutation effect generation is fundamentally an evidence-grounded reasoning task rather than a pure captioning problem. The model must integrate heterogeneous mutation signals, including local structural perturbations, distal contact-mediated propagation, and biochemical property constraints, into coherent functional descriptions. A reasoning-oriented decoder is therefore aligned with the structured conditioning objective of RipplePLM, where generation depends on combining multiple latent evidence pathways rather than decoding from a single undifferentiated representation.

Unless otherwise stated, all ESM-2 parameters are frozen, and the trainable components include the MPF/DDCA modules, PLChain auxiliary heads, the token projection layers, and LoRA adapters in the LLM. All experiments are conducted on a cluster with 8 NVIDIA A100 GPUs, each with 80GB memory. Training uses DeepSpeed ZeRO-2~\cite{rajbhandari2020zero}, bfloat16 mixed precision for memory efficiency.

\paragraph{Training hyperparameters.}
RipplePLM is trained in two stages. Stage~1 performs lightweight adaptation of the mutation-conditioning interface while keeping the LLM update small. In this stage, only the MPF module is trained. We train for 1 epoch on both temporal and structural splits using batch size 6 per GPU, learning rate $2{\times}10^{-4}$, and no property supervision. Stage~2 performs full mutation-aware adaptation with PLChain enabled. In this stage, LoRA adapters are applied to all attention and feed-forward projections (\texttt{q\_proj}, \texttt{k\_proj}, \texttt{v\_proj}, \texttt{o\_proj}, \texttt{up\_proj}, \texttt{gate\_proj}, \texttt{down\_proj}) with rank $r{=}64$ and $\alpha{=}128$. We train for 10 epochs on the temporal split and 2 epochs on the structural split using batch size 4 per GPU, gradient accumulation 2, and learning rate $5{\times}10^{-5}$. PLChain supervision is enabled only in Stage~2 through the thermostability and pH auxiliary heads.

Both stages use AdamW~\cite{loshchilov2017adamw} with 5\% linear warmup, bfloat16 mixed precision, and DeepSpeed ZeRO-2~\cite{rajbhandari2020zero}. The maximum sequence length is 1024. We use epoch-level checkpointing and retain the best Stage~2 checkpoint according to validation BLEU-4. The effective batch sizes are 48 in Stage~1 and 64 in Stage~2. Training is conducted on 8 NVIDIA A100 GPUs (80GB each). Stage~1 takes approximately 1.2h, while Stage~2 takes approximately 13.3h on the temporal split and approximately 2.7h on the structural split.

\paragraph{Pathway overlap regularization and full training objectives.}
\label{app:overlap}
For each example, we penalize overlap between the normalized direct and distal structural pathway weights:
\begin{equation}
    \mathcal{L}_{\text{overlap}} = \frac{1}{H}\sum_{h=1}^{H}\sum_{j=1}^{L} w^{\text{dir}}_j\,w^{\text{dist}}_{h,j}.
    \label{eq:pathway_overlap}
\end{equation}
The loss is averaged over the minibatch. It acts on the contact-derived weights used to construct the pathway seeds and attention biases. With $\lambda_{\text{overlap}}{=}0.05$ fixed throughout both stages, the complete objectives are
\begin{align}
    \mathcal{L}_{\text{stage 1}} &= \mathcal{L}_{\text{LM}} + \lambda_{\text{overlap}}\mathcal{L}_{\text{overlap}}, \\
    \mathcal{L}_{\text{stage 2}}(t) &= \mathcal{L}_{\text{task}}(t) + \lambda_{\text{overlap}}\mathcal{L}_{\text{overlap}},
\end{align}
where $\mathcal{L}_{\text{task}}(t)$ is the generation--property objective in Section~\ref{sec:training}.

\paragraph{Structural pathway-weight diagnostics.}
\label{app:pathway_diagnostics}
Figure~\ref{fig:diagnostics}b summarizes the structural weights $\mathbf{w}^{\text{dir}}$ and $\mathbf{w}^{\text{dist}}_h$ from DDCA. We analyze 110 temporal-test mutation examples with protein lengths of 100--300 residues and at least three predicted contacts above 0.5 at the mutation site. These criteria limit protein-size variation and ensure a sufficiently populated direct-contact neighborhood. Hop distances are shortest-path distances on the contact graph thresholded at 0.3. Distal weights are first averaged across heads. For each example and pathway, weights are summed by hop distance, restricted to reachable residues at hops 1--10, and renormalized over this range; we then report the mean and standard error across examples. This diagnostic characterizes the spatial coverage of the model's structural priors.

\paragraph{Inference.}
At inference time, the wild-type and mutant sequences are first encoded by frozen ESM-2 to construct the Mutation Perturbation Field and the predicted contact map used by DDCA. Generation uses greedy decoding with maximum output length 1024. PLChain auxiliary losses are disabled during inference, while the learned latent property tokens remain part of the conditioning sequence. Inference is performed on a single NVIDIA A100 GPU with batch size 1. The average end-to-end latency is approximately 1s per mutation, including ESM-2 encoding, DDCA computation, and autoregressive generation.

\paragraph{Compute usage.}
The reported experiments were conducted on an internal GPU cluster using 8 NVIDIA A100 GPUs (80GB memory each). The main reported runs correspond to one Stage~1 training run and one Stage~2 training run for each split configuration. Based on the measured runtimes, the total training cost for the experiments reported in the paper is approximately 150 GPU-hours. Additional compute was used for preliminary hyperparameter exploration, ablation studies, debugging, and failed runs that are not individually reported in the paper, bringing the estimated total project compute usage to approximately 600 GPU-hours. No large-scale pretraining was performed; all experiments rely on parameter-efficient fine-tuning with frozen ESM-2 representations and LoRA adaptation of the language model.

\paragraph{Dataset.}
We evaluate RipplePLM on MutaDescribe~\cite{mutaplm}, a mutation effect generation benchmark constructed from literature-derived mutation annotations. Each example contains a wild-type protein sequence, a point mutation, optional textual context, and a natural-language description of the mutation's functional consequence. We follow the official temporal and structural splits. The temporal split evaluates generalization to mutation descriptions from later literature, while the structural split groups test proteins according to their maximum sequence homology to training proteins, computed by MMseqs2. We use the provided mutation-effect descriptions as generation targets and retain the associated protein sequences and mutation identifiers for constructing wild-type/mutant inputs. When property labels are required by PLChain, we use the corresponding pseudo-label columns generated by PPT-Stab and EpHod for thermostability and pH optimum supervision. Detailed dataset construction and split definitions are provided in the original MutaDescribe paper~\cite{mutaplm}.

\paragraph{Prompt format.}
RipplePLM is trained in a chat-style instruction-following format. Table~\ref{tab:prompt_format} shows the prompt template used in our experiments. Here, \texttt{\{wt\}} and \texttt{\{mt\}} denote the wild-type and mutant amino acids, respectively, and \texttt{\{pos\}} denotes the 1-indexed mutation position in the protein sequence. The field \texttt{\{function description\}} is the provided wild-type protein function annotation, and \texttt{\{mutation-effect description\}} is the target natural-language description of the mutation's functional consequence.

\begin{table}[t]
\centering
\small
\setlength{\tabcolsep}{4pt}
\renewcommand{\arraystretch}{1.15}
\caption{Prompt template used by RipplePLM. The protein and property placeholders are replaced by learned latent tokens before being passed to the LLM.}
\label{tab:prompt_format}
\begin{tabularx}{\linewidth}{p{2.1cm} X}
\toprule
\textbf{Field} & \textbf{Content} \\
\midrule
System &
\texttt{You are an expert biochemist. Your task is to analyze a protein mutation.} \\
\midrule
User instruction & \texttt{Wild-type protein function: \{function description\}}
\texttt{Next is a feature of the mutation \{wt\} to \{mt\} at position \{pos\}. Please generate a \{brief summary / long detailed\} text to describe it. The feature is <|protein\_token|><|protein\_token|><|protein\_token|><|protein\_token|> <|thermo\_token|> <|ph\_token|>.} \\
\midrule
Assistant target &
\texttt{\{mutation-effect description\}} \\
\bottomrule
\end{tabularx}
\end{table}

During training and inference, the four repeated \texttt{<|protein\_token|>} placeholders are replaced by MPF/DDCA structural tokens derived from the wild-type/mutant perturbation field. The \texttt{<|thermo\_token|>} and \texttt{<|ph\_token|>} placeholders are replaced by PLChain latent property-query tokens for thermostability and pH optimum. The choice between \texttt{brief summary} and \texttt{long detailed} follows the availability of GPT-augmented descriptions in the training data.

We use the wild-type protein function description as input context rather than asking the model to generate it as part of an intermediate reasoning chain. This choice also ensures a fair matched-backbone comparison with MutaPLM$^{*}$. MutaPLM reports that replacing model-generated function descriptions with ground-truth function descriptions improves performance; therefore, in our MutaPLM$^{*}$ reimplementation, we also provide the ground-truth function description as prompt context. This keeps the comparison focused on the mutation-conditioning mechanism rather than on differences in access to functional context.

\section{Low-N Fitness Regression}
\label{app:fitness}

We follow the low-$N$ fitness regression protocol of MutaPLM~\cite{mutaplm}.
Given only 192 labeled single-point mutations, we train a lightweight regression head on frozen mutation features and evaluate Spearman correlation over five random runs.
For fairness, RipplePLM uses the same frozen ESM-2 feature views as the baseline setting and augments them with MPF features extracted without updating the generator.

RipplePLM obtains the highest mean Spearman correlation on both datasets, although the standard deviations overlap with the strongest baselines.
We therefore view this experiment as auxiliary evidence that MPF representations retain fitness-relevant information, rather than as a definitive claim of superiority on quantitative prediction.

\paragraph{Datasets.}
We evaluate on two ProteinGym v1.3 datasets:
(1)~{Spike-ACE2}: 3802 single-point mutations, wild-type length 1273 aa;
(2)~{avGFP}: 1084 single-point mutations filtered from 51714 entries, wild-type length 238 aa.

\paragraph{Training.}
The regression head is trained with 192 training samples and 48 validation samples using mean-squared error.
Spearman correlation is computed on the validation set.

\section{Full Structural Split Results}
\label{app:structural_full}

Table~\ref{tab:structural_full} reports the full six-metric results for RipplePLM on the structural split.

\begin{table}[ht]
\centering
\caption{Full RipplePLM results on the structural split across all six metrics.}
\label{tab:structural_full}
\begin{tabular}{@{}lcccccc@{}}
\toprule
Level & BLEU-2 & BLEU-4 & METEOR & ROUGE-1 & ROUGE-2 & ROUGE-L \\
\midrule
Easy   & 22.57 & 16.07 & 37.51 & 45.26 & 31.75 & 42.91 \\
Medium & 12.67 &  5.70 & 24.72 & 34.65 & 17.27 & 31.56 \\
Hard   & 11.91 &  5.73 & 24.30 & 34.05 & 17.05 & 31.27 \\
\midrule
Average & 15.78 & 9.36 & 29.30 & 38.37 & 22.54 & 35.65 \\
\bottomrule
\end{tabular}
\end{table}

\section{Case Studies}
\label{app:case}

To qualitatively analyze model behavior, we compare RipplePLM with MutaPLM* on representative examples from the Easy, Medium, and Hard subsets of the structural split. We select cases that cover partial matches, quantitative effects, interaction-specific descriptions, and competitive counterexamples.

\begin{table*}[t]
\centering
\small
\setlength{\tabcolsep}{3pt}
\renewcommand{\arraystretch}{1.15}
\caption{
Representative case studies on the structural split of MutaDescribe.
}
\label{tab:case_study}
\begin{tabularx}{\textwidth}{p{1.2cm} |X| X| X}
\toprule
\textbf{Mutation} & \textbf{Reference} & \textbf{RipplePLM} & \textbf{MutaPLM*} \\
\midrule
\textbf{N226A}
&
100-fold decrease in GTPase activity. 5-fold decrease of affinity for GTP.
&
100-fold decrease in GTPase activity.
&
Abolishes GTP hydrolysis.
\\
\midrule
\textbf{G93A}
&
Compromised interaction with TDR.
&
Compromised interaction with TDR. The mutation in TDIF reduces its binding affinity to TDR.
&
Reduced ability to repress tracheary element differentiation.
\\
\midrule
\textbf{Q100L}
&
Probable constitutively active mutant unable to hydrolyze GTP; increases degradation of the transferrin receptor.
&
Probable constitutively active mutant unable to hydrolyze GTP; localizes to the cytoplasm.
&
In a breast cancer sample; somatic mutation.
\\
\midrule
\textbf{A715R}
&
Increased ability to package the SNARE SEC22B cargo into COPII vesicles. Has no effect on other cargos packaged into COPII vesicles.
&
Increased ability to bind and package the SNARE SEC22B into transport vesicles.
&
Enables interaction with SEC31A.
\\
\midrule
\textbf{Q49A}
&
Reduces ATP binding and ATPase activity in vitro.
&
Reduces RNA binding affinity. However, no specific functional changes are described for this mutation.
&
Loss of ATPase activity.
\\
\bottomrule
\end{tabularx}

\end{table*}

These cases show that RipplePLM is most effective when the reference description depends on mutation-specific structural or mechanistic evidence, such as GTPase reduction, TDIF--TDR interaction, or SEC22B cargo packaging. The examples also reveal realistic error modes: RipplePLM can omit secondary clauses, substitute downstream outcomes, or confuse related biochemical activities. The competitive Q49A case further indicates that structured conditioning improves many mutation-specific descriptions but does not eliminate all enzyme-activity errors.

\section{Additional Generation Evaluation}
\label{app:additional_evaluation}

\subsection{Biology-Aware Evaluation}
\label{app:expert_eval}

Following the four-category framework in MutaPLM~\cite{mutaplm} (Appendix C.3, Table A8), a PhD-level biology expert evaluated all 1,248 RipplePLM outputs on the structural test set against their reference descriptions. Each output was assigned to Accurate, Relevant, Opposite, or Irrelevant based on agreement in the affected biological function or process and the direction of the mutation effect. Table~\ref{tab:expert_eval} reports the expert-reviewed distribution alongside the published MutaPLM results. RipplePLM achieves an Accurate-or-Relevant rate of 48.56\%, an increase of 8.50 percentage points, while the Opposite rate decreases from 7.13\% to 6.41\%. This evaluation complements the lexical metrics with an assessment of biological fact agreement.

\begin{table}[htbp]
\centering
\small
\caption{Biology-aware evaluation on the structural split. MutaPLM results are reported in the original paper~\cite{mutaplm}; RipplePLM results are based on expert review of 1,248 outputs. Values are percentages.}
\label{tab:expert_eval}
\begin{tabular}{@{}lrr@{}}
\toprule
Category & MutaPLM & RipplePLM \\
\midrule
Accurate $\uparrow$ & 17.47 & 21.55 \\
Relevant $\uparrow$ & 22.60 & 27.00 \\
Accurate + Relevant $\uparrow$ & 40.06 & 48.56 \\
Opposite $\downarrow$ & 7.13 & 6.41 \\
Irrelevant $\downarrow$ & 52.80 & 45.03 \\
\bottomrule
\end{tabular}
\end{table}

\subsection{Response Length}
\label{app:length_control}

RipplePLM generates an average of 30.65 words per description on the temporal split, compared with 28.03 words in the references. On the structural split, its descriptions are shorter than the references (21.73 vs.\ 31.87 words). To assess performance under a reference-length constraint, we truncated the 732 temporal outputs that exceeded their reference lengths, leaving the remaining outputs unchanged, and rescored all 1,611 examples using the same tokenizer and evaluation metrics.

As shown in Table~\ref{tab:length_control}, truncation increases ROUGE-L from 28.56 to 32.68 and also improves METEOR, ROUGE-1, and ROUGE-2. BLEU-2 and BLEU-4 decrease slightly, while all six scores remain above the published MutaPLM results. Output length is negatively correlated with per-example ROUGE-L on both splits (Pearson $r\approx-0.32$ temporal and $-0.24$ structural). These results indicate that additional output length is unlikely to be the main source of the generation gains.

\begin{table}[htbp]
\centering
\small
\caption{Temporal-split performance before and after truncating over-length RipplePLM outputs to their reference lengths. B/R denote BLEU/ROUGE. The published MutaPLM scores are included for comparison.}
\label{tab:length_control}
\begin{tabular}{@{}lrrrrrr@{}}
\toprule
Model / output setting & B-2 & B-4 & METEOR & R-1 & R-2 & R-L \\
\midrule
MutaPLM & 10.83 & 6.15 & 17.84 & 18.99 & 6.92 & 16.51 \\
RipplePLM & 13.41 & 6.68 & 25.23 & 31.15 & 16.00 & 28.56 \\
RipplePLM (truncated) & 12.58 & 6.58 & 25.59 & 35.33 & 18.64 & 32.68 \\
\bottomrule
\end{tabular}
\end{table}

\subsection{Statistical Uncertainty}
\label{app:bootstrap}

We estimate 95\% bootstrap confidence intervals for RipplePLM by resampling test examples 10,000 times within each split. Table~\ref{tab:bootstrap_ci} reports intervals for BLEU-2 and ROUGE-L on the temporal ($n=1,611$) and structural ($n=1,248$) splits. These intervals characterize sampling uncertainty over test examples. Their lower bounds exceed the corresponding published MutaPLM point estimates on both splits.

\begin{table}[htbp]
\centering
\small
\caption{RipplePLM scores with 95\% bootstrap confidence intervals.}
\label{tab:bootstrap_ci}
\begin{tabular}{@{}lcc@{}}
\toprule
Metric & Temporal & Structural \\
\midrule
BLEU-2 & 13.41 [12.43, 14.43] & 15.78 [14.42, 17.15] \\
ROUGE-L & 28.56 [27.51, 29.63] & 35.65 [34.31, 37.01] \\
\bottomrule
\end{tabular}
\end{table}

\subsection{Recent General-Purpose LLMs}
\label{app:recent_llms}

We evaluate GPT-5.4, DeepSeek-V4-Flash, and DeepSeek-V4-Pro on all 1,248 structural-test examples in a one-shot setting without external tools or retrieval. Table~\ref{tab:recent_llms} compares their results with the reported GPT-4-0613 one-shot baseline and RipplePLM. The recent LLMs achieve broadly comparable scores to GPT-4-0613, while RipplePLM retains a substantial advantage on both metrics.

\begin{table}[htbp]
\centering
\small
\caption{Recent general-purpose LLMs on the structural split. Prompted LLMs use one-shot evaluation; GPT-4-0613 results are from MutaPLM~\cite{mutaplm}.}
\label{tab:recent_llms}
\begin{tabular}{@{}lrr@{}}
\toprule
Model & BLEU-2 & ROUGE-L \\
\midrule
GPT-4-0613 & 3.00 & 9.14 \\
GPT-5.4 & 2.07 & 8.49 \\
DeepSeek-V4-Flash & 3.13 & 10.13 \\
DeepSeek-V4-Pro & 3.31 & 9.96 \\
RipplePLM & 15.78 & 35.65 \\
\bottomrule
\end{tabular}
\end{table}

\section{Additional Model and Input Analyses}
\label{app:additional_controls}

\subsection{Wild-Type Function Context}
\label{app:function_context}

RipplePLM and MutaPLM$^{*}$ receive the same ground-truth wild-type function description in the matched-backbone comparison. To assess the contribution of this input, we remove the function description from RipplePLM on the structural split and compare with the function ablation reported in Table 4 of MutaPLM~\cite{mutaplm}. Table~\ref{tab:function_context} reproduces both MutaPLM conditions from that ablation, whose with-function score differs from the main benchmark result in Table~\ref{tab:structural}.

Removing function context reduces RipplePLM's ROUGE-L by 7.10 points (19.9\%), compared with a 5.26-point decrease (22.1\%) for MutaPLM. RipplePLM retains 80.1\% of its score, exceeds no-function MutaPLM by 10.01 points, and exceeds MutaPLM with the ground-truth function description by 4.75 points. The function descriptions themselves have a ROUGE-L overlap of 9.76 with the mutation-effect references, and no reference is an exact substring of its function description. Function context provides useful information, while the removal control shows that RipplePLM retains its comparative advantage without this input.

\begin{table}[htbp]
\centering
\small
\caption{Effect of wild-type function descriptions on structural-split ROUGE-L. MutaPLM values are from its published function ablation~\cite{mutaplm}, Table 4.}
\label{tab:function_context}
\begin{tabular}{@{}lrrrr@{}}
\toprule
Model & With function & Without function & Drop & Relative drop \\
\midrule
MutaPLM & 23.80 & 18.54 & 5.26 & 22.1\% \\
RipplePLM & 35.65 & 28.55 & 7.10 & 19.9\% \\
\bottomrule
\end{tabular}
\end{table}

\subsection{Protein Encoder Size}
\label{app:encoder_size}

We replace the frozen ESM-2 150M encoder with ESM-2 650M, adjusting only the MPF input projection to accommodate the embedding dimension and retaining the other hyperparameters and training settings. The larger encoder improves four of six structural-split metrics (Table~\ref{tab:encoder_size}), including METEOR (29.30 to 30.45) and ROUGE-L (35.65 to 36.43), with small decreases in BLEU-2 and BLEU-4. This comparison suggests potential benefits from greater encoder capacity under the existing training configuration. We use the 150M encoder for the main experiments to balance predictive performance and computational cost.

\begin{table}[htbp]
\centering
\small
\caption{Effect of frozen ESM-2 encoder size on the structural split. B/R denote BLEU/ROUGE.}
\label{tab:encoder_size}
\begin{tabular}{@{}lrrrrrr@{}}
\toprule
Encoder & B-2 & B-4 & METEOR & R-1 & R-2 & R-L \\
\midrule
ESM-2 150M & 15.78 & 9.36 & 29.30 & 38.37 & 22.54 & 35.65 \\
ESM-2 650M & 15.12 & 9.22 & 30.45 & 38.84 & 23.63 & 36.43 \\
\bottomrule
\end{tabular}
\end{table}

\subsection{Predicted-Contact Confidence}
\label{app:contact_confidence}

We summarize model-reported contact confidence as the mean ESM-2 150M contact probability over the top-$L$ long-range residue pairs ($i<j$, $j-i\geq24$), where $L$ is the wild-type sequence length. When fewer than $L$ eligible pairs are available, we average over all of them. Prior work on attention-based contact prediction motivates the use of these probabilities as a confidence statistic~\cite{rao2021contact}. Of the 1,248 structural-test examples, 1,247 have eligible long-range contacts; we divide these examples into three approximately equal-sized groups by confidence and apply the expert-reviewed labels from Appendix~\ref{app:expert_eval}.

Table~\ref{tab:contact_confidence} shows that the Accurate-or-Relevant rate increases from 44.23\% in the lower-confidence group to 54.22\% in the higher-confidence group. This analysis measures the association between generation performance and model-reported contact confidence, rather than contact accuracy against experimentally determined structures. The pattern is consistent with the use of predicted contacts as a structural prior, with content-based attention contributing alongside that prior.

\begin{table}[htbp]
\centering
\small
\caption{Biology-aware evaluation across predicted-contact confidence groups on the structural split. Category values are percentages.}
\label{tab:contact_confidence}
\begin{tabular}{@{}lrrrr@{}}
\toprule
Confidence group & $N$ & Accurate & Relevant & Accurate + Relevant \\
\midrule
Lower third & 416 & 19.47 & 24.76 & 44.23 \\
Middle third & 416 & 19.71 & 27.40 & 47.12 \\
Higher third & 415 & 25.54 & 28.67 & 54.22 \\
\bottomrule
\end{tabular}
\end{table}


\clearpage

\end{document}